\documentclass[aps,prl,twocolumn,epsf,epsfig,showpacs]{revtex4}
\usepackage{graphicx}
\begin{document}
\newcommand{\be}{\begin{equation}}
\newcommand{\ee}{\end{equation}}
\newcommand{\beqa}{\begin{eqnarray}}
\newcommand{\eeqa}{\end{eqnarray}}
\def\nn{\nonumber}
\def\l({\left(}
\def\r){\right)}

\title{A method of measurement of extragalactic magnetic fields by TeV gamma ray telescopes}

\author{A.~Neronov}
\affiliation{INTEGRAL Science Data Center, Chemin d'\'Ecogia 16, 1290 Versoix,
  Switzerland and Geneva Observatory, 51 ch. des Maillettes, CH-1290 Sauverny,
  Switzerland}
\author{D.V.~Semikoz}
\affiliation{APC, Coll\`ege de France, 
11, pl. Marcelin Berthelot, Paris 75005, France}
\affiliation{Institute for Nuclear Research RAS, 60th October Anniversary prosp. 7a,
Moscow, 117312, Russia}

\begin{abstract} We propose a method of measurement of extragalactic magnetic
fields  in  observations of TeV $\gamma$-rays from distant sources.  Multi-TeV
$\gamma$-rays  from these sources interact with the infrared photon background
producing secondary electrons and positrons, which can
be just slightly  deflected by extragalactic magnetic fields before they emit
secondary $\gamma$ rays via inverse  Compton
scattering of cosmic microwave background photons. Secondary $\gamma$-rays emitted
toward an observer on the Earth can be detected as extended
emission around initially point source. Energy dependent angular profile of
extended emission is determined by the characteristics  of extragalactic magnetic field along
the line of sight. Small magnetic fields  $ B \le 10^{-12}$ G in the voids of the
large scale structure can be measured in this way.
\end{abstract}


\maketitle

The problem of the origin of $1-10\ \mu$G magnetic fields in galaxies  and galaxy
clusters is one of the long standing problems of astrophysics/cosmology~(see
\cite{primordial_review} for a review).  Such fields are thought to be
produced  either via dynamo mechanism \cite{dynamo} or via compression of
primordial magnetic field during the large scale structure (LSS) formation
\cite{primordial_review}. In either of these two scenarios one assumes the
existence of a small "seed" primordial magnetic field of cosmological origin.
The assumed strength of the primordial magnetic field and its  correlation
length are strongly model dependent.  Moderate experimental limit on
the present day strength of primordial magnetic field $B<10^{-9}$G comes from the
limit on rotation measure of emission from distant quasars, assuming (largely
uncertain) correlation length $l_c\sim 1$~Mpc~\cite{kronberg}. 
Similar restriction comes from the analysis of anisotropies 
of  the cosmic microwave background (CMB)~\cite{Barrow:1997mj}.

Recently a significant progress was achieved in calculations of the
tree-dimensional structure of extragalactic  magnetic fields (EGMF)
~\cite{defl_Sigl,defl_Dolag}. These  calculations  are based on the numerical
simulations of LSS  formation. The  main result of  the magnetic field
simulations is that, similarly to the LSS, the magnetic field structure is
filamentary, with large voids, filaments  and knots. The strength of the
primordial magnetic field in these simulations is chosen in such a way as to
reproduce the measured fields $B\sim 1\ \mu$G  in Galaxy clusters. New
simulations allow to estimate the cumulative volume filling factor ${\cal
V}(B)$ for EGMF with the strength close to a typical value $B$.  One should
note that the results of simulations by different groups differ dramatically.
In particular, in  Ref.~\cite{defl_Sigl} it is claimed that  ${\cal
V}(B>10^{-9}\mbox{ G})\simeq 0.15$, while ${\cal V}(B<10^{-12}\mbox{ G})\simeq
0.07$.  To the contrary, in Ref.~\cite{defl_Dolag}  ${\cal V}(B>10^{-9}\mbox{
G})\simeq 2\times10^{-4}$  while ${\cal V}(B<10^{-12}\mbox{ G})\simeq 0.7$. 
Recent independent calculation of  ref.~\cite{Bruggen2005} suggests that 
$B_{\rm } > 10^{-9} $ G fields fill $\sim 1$\% of volume, which is somewhat in
between of the results of Ref.~\cite{defl_Sigl} and Ref.~\cite{defl_Dolag}.

Obviously, the best way to resolve the above controversy of numerical
calculations is to measure the EGMF outside the galaxies and galaxy clusters. 
In this paper we propose a way of measurement of EGMF which is
sensitive to the very weak magnetic fields $B\le 10^{-12}$~G and, therefore,
could be used to probe, for the first time, the magnetic fields in the voids of
the LSS.  

The idea is to look for the extended TeV $\gamma$-ray emission around distant
sources. It is known that multi-TeV $\gamma$-rays from these
sources could not reach the Earth directly because of the pair production on
the extragalactic (infrared) background light (EBL). 
The pair production on EBL should lead to the exponential 
suppression of the $\gamma$-ray flux from the source,
\begin{equation} 
\label{absorb}
F(E_{\gamma_0})=F_0(E_{\gamma_0})\exp\left[-\tau(E_{\gamma_0},z)\right]. 
\end{equation} 
Here $F(E_{\gamma_0})$ is the detected spectrum,
$F_0(E_{\gamma_0})$ is the initial spectrum of the source and $\tau(E_{\gamma_0},z)$ 
is the optical depth with respect to the pair production
on EBL, which is a function of the primary photon energy $E_{\gamma_0}$ and of the
redshift of the source $z$ \cite{aharon}
\begin{equation}
\tau =5\div 10 \left[\frac{z}{0.1}\right]\left[
\frac{\rho_{IR/O}}{(5\div 10) \mbox{ nW/(m}^2\mbox{sr)}}\right]
\left[\frac{E_{\gamma_0}}{10\mbox{ TeV}}\right]
\label{tau}  
\end{equation}
(we allow for an uncertainty of about a factor of 2 in the density of the 
infrared/optical background, $\rho_{IR/O}$, taking into account the recent 
controversy in the measurements of COBE/DIRBE and the upper limits on
$\rho_{IR/O}$ imposed by the TeV observations of distant blazars \cite{ebl}).  

The $e^+e^-$ pairs of the energy $E_e$ produced in interactions of multi-TeV
$\gamma$-rays  with EBL photons produce secondary $\gamma$-rays via inverse
Compton scattering (ICS) of the CMB photons  to the energies 
\begin{equation}
E_{\gamma}=\frac{4}{3}\epsilon_{CMB}\frac{E_e^2}{m_e^2}\simeq 1.2 
\left[\frac{E_{\gamma_0}}{40 \mbox{ TeV}}\right]^2\mbox{ TeV}
\label{Esec}
\end{equation}
(Here $\epsilon_{CMB}=6\times 10^{-4}$~eV is the typical energy of CMB photons. 
We also have assumed that the energy of primary $\gamma$-ray is
$E_{\gamma_0}\simeq 2E_e$). Upscattering of the infared/optical backgound photons gives
sub-dominant contribution to the ICS because 
the energy density of CMB, $\rho_{CMB}\sim 10\div 100\rho_{IR/O}$, 
is much higher than the
density of the infrared/optical background. 

It  is conventionally assumed that as soon as  trajectories of electrons are
deflected by the EGMF, the secondary photons are not emitted in the direction
toward the Earth and do not reach the telescope. However, the above way of
reasoning is correct only if one calculates only the reduction of the flux of
photons emitted initially exactly in the direction of observer. Deflections of
electrons produced by the $\gamma$-rays which were initially emitted slightly
away from the observer, could lead to "redirection" of the secondary   cascade
photons toward the observer.  This effect leads to the appearance of additional
"cascade" contribution to the source flux. 

\begin{figure}
\includegraphics[width=\columnwidth]{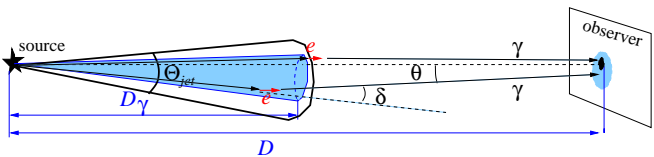}
\caption{Schematic representation of formation of extended TeV emission around a
distant point source due to the cascade on the EBL in small EGMF.}
\label{fig:cartoon}
\end{figure}

The cascade contribution can be distinguished from the "primary" source
contribution with the help of imaging capabilities of the TeV Cherenkov
telescopes. The secondary cascade $\gamma$-rays are emitted  at large distances
from the source,  from an extended  conical region (blue shaded region in  Fig.
\ref{fig:cartoon}). The opening angle of the cone is determined by the
deflection angle of secondary electrons, $\delta$,  its  vertex coincides with
the point source,  and its height is equal to the mean free path 
$D_{\gamma_0}=D/\tau\simeq 10\div 20 \left[E_{\gamma_0}/40\mbox{
TeV}\right]^{-1}$~Mpc ($D$ is the distance to the source) of the multi-TeV
photons through  the EBL.   Since the line of sight is located inside of the
cone, the cone appears as an extended emission around the initial point source
(Fig. \ref{fig:cartoon}). 

The mean  free path of the secondary electrons with respect to the ICS of CMB
photons, $\lambda_e=(\sigma_Tn_{CMB})^{-1} \simeq 1$~kpc ($\sigma_T=6.7\times
10^{-25}\mbox{cm}^2$ is the Thompson cross-section  and
$n_{CMB}=411/\mbox{cm}^3$ is CMB photon density), is much shorter than
$D_{\gamma_0}$ and one can assume that electrons immediately re-emit
$\gamma$-rays via the ICS. The deflection angle of electrons, $\delta$,  is
determined by the ratio of $\lambda_e$ to the Larmor radius in the EGMF of the
strength $B$, $r_L\simeq 6.6\times 10^{22}\left[E_e/20\mbox{ TeV}\right]
\left[10^{-12}\mbox{ G}/B\right]$~cm.  Assuming $\lambda_e\ll r_L$ one can
find  $\delta\simeq\lambda_e/r_L=3.2^\circ  \left[20
~\mbox{TeV}/E_e\right]\left[B/10^{-12}\mbox{ G}\right]$.   A simple geometrical
calculation shows that  the angular size of extended emission produced via
redirection of cascade photons toward the Earth is (see Fig.
\ref{fig:cartoon}) 
\begin{equation}
\theta_{ext}\simeq \frac{D_{\gamma_0}}{D}\delta =
\frac{0.3^\circ}{\tau(E_{\gamma_0},z)}\left[\frac{1
~\mbox{TeV}}{E_{\gamma}}\right]^{1/2}\left[\frac{B}{10^{-13}\mbox{ G}}\right]
\label{thetahalo}
\end{equation}
(throughout the paper we use the relation between the primary photon energy, 
$E_{\gamma_0}$, the energy of secondary electron, $E_e$ and of the secondary
photon, $E_\gamma$ given by Eq. (\ref{Esec})).

Comparing the distance covered by the photons which come directly from the
point source with the distance covered by (primary and secondary) cascade
photons,   one can find (from the same geometrical calculation) the typical
time delay of  the secondary photons, $t_d\simeq \delta^2 D(\tau-1)/2\tau^2$.
Contrary to the point source flux, the flux in the cascade component is not
variable, since it  is equal to the point source flux averaged over a large
time scale $t_d$ (e.g.  for $\delta\sim 5^\circ$,
$z=0.03$ and $E_{\gamma_0} \sim 40$ TeV, one finds $t_d \sim 10^5$~yr).

Since each secondary cascade photon carries only a small fraction of the
primary photon energy (see Eq. (\ref{Esec})), the flux of the extended source,
integrated over the region of the angular size $\theta_{ext}$  is a small
fraction  $\alpha\simeq2E_\gamma/E_{\gamma_0}\simeq 
0.06\left[E_{\gamma_0}/40\mbox{ TeV}\right]$ of initial point source flux
$\left<F_0(E_{\gamma_0})\right>$ (angle brackets signify averaging of the point source
flux over the time scale $t_d$). 
The extended source flux is suppressed at high energies 
because  the secondary cascade photons can  themselves produce pairs in
interactions with  EBL photons. The suppression factor is
$\exp\left(-\tau(E_\gamma,z)\right)$ where $\tau$ is given by (\ref{tau}).  
Apart from the absorption at high energies, the spectrum of the extended source
is modified  if only a fraction of the mean free path of the primary
$\gamma$-ray, $D_{\gamma_0}$,  goes through the voids with small magnetic 
field $B$. This fraction can be expressed through the volume filling factor of 
${\cal V}(B)$, as $D_{\rm voids}/D_{\gamma_0}\simeq {\cal V}^{1/3}$. 
Combining all the correction/suppression factors   one can find a relation
between the average absorbed point source flux $\left<F(E_{\gamma_0})\right>$ and
the flux of extended source integrated over the region $\theta<\theta_{ext}$:
\begin{equation}  F_{ext}(E_\gamma,\theta_{ext}) \simeq  
0.06{\cal V}^{1/3} \frac{\left(e^{\tau(E_{\gamma_0},z)}-1\right)}{e^{\tau(E_\gamma,z)}} 
\left[\frac{E_{\gamma}}{1\mbox{ TeV}}\right]^{1/2}
\left<F(E_{\gamma_0})\right> 
\label{sext1}    
\end{equation} 

\begin{figure}
\includegraphics[width=0.8\columnwidth]{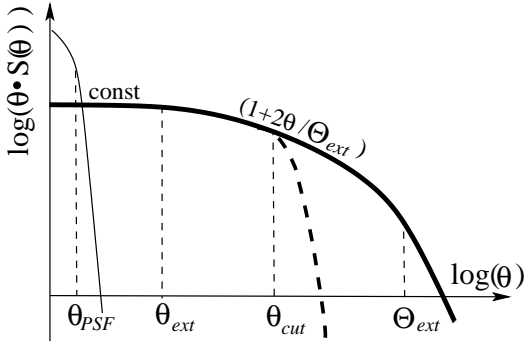}
\caption{Surface brightness profile of Extended source produce by the cascade on
EBL.}
\label{fig:sb}
\end{figure}

Cascade electrons loose about half of their energy after upscattering of many CMB photons
over the energy attenuation distance  $\Lambda_e \sim
(E_e/E_\gamma)\lambda_e\sim 20\left[20\mbox{ TeV}/ E_e\right]\lambda_e$.  ICS
emission produced by these electrons over the distance $\Lambda_e>\lambda_e$
forms a powerlaw tail around the core of the  extended source,
$\theta<\theta_{ext}$, as it is shown in Fig. \ref{fig:sb}.  Since for electron
propagtion distances $d<\Lambda_e$ the deflection angle 
scales linearly with $d$, the surface brightness profile  of the tail can be
found by integrating the ICS energy loss along electron trajectory,
${\cal S}(\theta)\sim \theta^{-1}(1+2\theta/\Theta_{ext})$, where
$\Theta_{ext}$ is  the opening angle of a cone  into which the  ICS
photons produced by the cascade electrons over the energy attenuation length 
are emitted, $\Delta=\Lambda_e/r_L\simeq 6^\circ\left[20\mbox{
TeV}/E_e\right]^2\left[B/10^{-13}\mbox{ G}\right]$.  Using Fig.
\ref{fig:cartoon} one finds, 
\begin{equation}
\label{Thetaext}
\Theta_{ext}=\frac{D_{\gamma_0}}{D}\Delta\simeq \frac{6^\circ}{\tau(E_\gamma,z)}
\left[\frac{1\mbox{
TeV}}{E_\gamma}\right]\left[\frac{B}{10^{-13}\mbox{ G}}\right]
\end{equation}

The fraction of the total flux of extended source integrated over the region 
$\theta<\Theta_{ext}$ is about the initial point source flux $\left<
F_0(E_{\gamma_0})\right>$ modified at high energies by the effect of
absorption of the secondary $\gamma$-rays and by the volume-filling-dependent
factor ${\cal V}^{1/3}$,
\begin{equation}
F_{ext}(E_\gamma,\Theta_{ext})\simeq {\cal
V}^{1/3}\frac{\left(e^{\tau(E_{\gamma_0},z)}-1\right)}{e^{\tau(E_\gamma,z)}}
\left<F(E_{\gamma_0})\right>
\label{sext2}
\end{equation}
(compare with  Eq. (\ref{sext1})).

Up to now we have assumed that the primary $\gamma$-rays are emitted by the
point source isotropically. However,  most of the distant sources of TeV
$\gamma$-rays are blazars. The primary  $\gamma$-ray emission from blazars is
known to be concentrated along the direction of the jet with a typical opening
angle $\Theta_{jet}\sim  \Gamma_{jet}^{-1}\sim 5^\circ$ ($\Gamma_{jet}\sim 10$
is the typical bulk Lorentz factor of the jet). There is no difference between
isotropically and anisotropically emitting source till the development of the
cascade on EBL leads to redistribution of directions of secondary $\gamma$-rays
within a cone with  an  opening angle smaller than the openting angle of the
jet (see Fig. \ref{fig:cartoon}). However, as soon as the deflected cascade
photons are emitted into a cone with opening angle larger than $\Theta_{jet}$, 
the flux of extended source becomes suppressed.  This leads to a steepening of 
the surface brighness profile of the extended source at the angles
$\theta>\theta_{cut}$ where
\begin{equation}
\theta_{cut}\simeq \frac{D_{\gamma_0}}{D-D_{\gamma_0}}\Theta_{jet}\simeq\frac{\Theta_{jet}}{\tau-1}
\label{thetacut}
\end{equation}
 
Suppose that extended sources with characteristic (energy-dependent)  surface
brighness profiles shown in Fig. \ref{fig:sb} are found around a number of
distant point sources.  In this case measuring the extended source parameters
one can constrain  the characteristics of EGMF  along the source directions.
For example,  combining  Eqs. (\ref{thetahalo}) and (\ref{sext1}) one finds
that measurements of $\theta_{ext}(E_\gamma)$ and
$F_{ext}(E_\gamma,\theta_{ext})$  provides a measurement of the volume filling
factor ${\cal V}(B)$  of EGMFs of particular strength $B$: 
\begin{eqnarray}
&&B\simeq 
10^{-13}\tau(E_{\gamma_0},z)\left[\frac{E_{\gamma}}{1~\mbox{TeV}}\right]^{1/2} 
\left[\frac{\theta_{ext}}{0.5^\circ}
\right]\mbox{~G}\nonumber\\
&&{\cal V}(B)\simeq
1\left[\frac{F_{ext}(E_\gamma)e^{\tau(E_\gamma,z)}}{0.06\left<F(E_{\gamma_0})\right>
(e^{\tau(E_{\gamma_0},z)-1})}\right]^{3}
\left[\frac{1\mbox{
TeV}}{E_\gamma}\right]^{3/2}
\label{EB}
\end{eqnarray}
Othervise, from (\ref{Thetaext}), (\ref{sext2}) one finds that information about
$B$, ${\cal V}(B)$ can be extracted from the measurement of  
$\Theta_{ext}(E_\gamma),F_{ext}(E_\gamma,\Theta_{ext})$:
\begin{eqnarray}
&&B\simeq10^{-13}\tau(E_{\gamma_0},z)\left[\frac{E_\gamma}{1\mbox{~TeV}}
\right]\left[\frac{\Theta_{ext}}{6^\circ}\right]\mbox{~G}\nonumber\\
&&{\cal V}(B) \simeq 
\left[\frac{F_{ext}(E_\gamma,\Theta_{ext})e^{\tau(E_\gamma,z)}}{\left<F(E_{\gamma_0})\right>
(e^{\tau(E_{\gamma_0},z)}-1)}\right]^3
\label{EB1}
\end{eqnarray}
Determination of characteristics of EGMF from  either (\ref{EB}) or (\ref{EB1}) constitutes the essence of the proposed
method of measurement of EGMF.

Maximal EGMF strength which can be probed is 
found from observation that  $\theta_{ext}$ should not be larger than
$\theta_{cut}$ (to be measurable). 
Since $\theta_{ext}\sim E^{-1/2}$ (see (\ref{thetahalo})), the
maximal EGMF strength is determined by the condition that
$\theta_{ext}(E_{max})<\theta_{cut}$ where $E_{max}$ is 
the highest energy at which a
source is detected. Combining  (\ref{thetahalo}) and
(\ref{thetacut}) one finds
\begin{equation} 
B_{max}\simeq 3\times
10^{-12}\frac{\tau(E_{\gamma_0},z)}{\tau(E_{\gamma_0},z)-1}\left[
\frac{E_{max}}{10\mbox{ TeV}}\right]^{1/2}
\left[\frac{\Theta_{jet}}{5^\circ}\right]\mbox{ G}~, 
\label{Bmax} 
\end{equation}

The weakest magnetic fields which can be probed 
is found from observation that 
$\Theta_{ext}$ can not be measured if it becomes smaller than the size of the
point spread function of the telescope, $\theta_{PSF}$. 
Since $\Theta_{ext}\sim E_\gamma^{-1}$ (see (\ref{Thetaext})), the
largest source extension is achieved at lowest energies. 
Taking into account that for present generation instruments  the low energy
threshold is situated at the energies $E_{min}\sim 100$~GeV and that the
typical point spread function of a Cherenkov telescope has the size
$\theta_{PSF}\sim 0.1^\circ$, one finds from  (\ref{Thetaext})
and (\ref{EB1}) 
\begin{equation} 
B_{min}\simeq 10^{-16} \tau(E_{\gamma_0},z) \left[
\frac{E_{min}}{100\mbox{ GeV}}\right]
\left[\frac{\theta_{PSF}}{0.1^\circ}\right]\mbox{ G} 
\label{Bmin} 
\end{equation}

As an example of implementation of the proposed method of measurement of EGMF,
let  us find the constraints on ${\cal V}(B)$ imposed by the non-observation of
extended emission around Mkn 501 by HEGRA.  An upper limit on the extended
emission flux within $0.5^\circ$ around Mkn 501,
$F_{ext,0.5^\circ}<10^{-3}F_{flare}$ ($F_{flare}$ is the flux from Mkn 501 in
the flaring state) was derived by HEGRA collaboration \cite{mkn501halo}).
Unfortunately, neither the energy dependence of the derived upper limit nor 
the assumed surface brightness profile are reported in \cite{mkn501halo}. We assume in the
following that the restriction was put on the flux in the core of the extended
source, $\theta<\theta_{ext}$.  The quiescent flux of Mkn
501 is about 10 times lower than the flux in the flaring state which means that
HEGRA observation imposes a restriction on the extended emission flux at the
level of 
$F_{ext}(0.5\mbox{~TeV},\theta_{ext})< 0.01\left<F(0.5\mbox{ TeV}
)\right>$
The primary $\gamma$ rays
which produce extended emission at $E_\gamma\simeq 0.5$~TeV have energies
$E_{\gamma_0}\simeq 26$~TeV. Optical depth for such $\gamma$ rays is large,
$\tau(26$~TeV,$ 0.03)=4-8$ (see (\ref{tau}))  even for
the relatively nearby source  Mkn 501 ($z=0.03$).   The spectrum of Mkn 501 is
characterized by the photon index $\Gamma= 2.3\pm 0.2$ \cite{mkn501_spec},
which means that 
$\left<F_{0}(26\mbox{~TeV})\right>
\simeq 0.3
\left<F_0(0.5\mbox{~TeV})\right>$ or
\begin{equation}
\label{mkn3}
F_{ext}(E_\gamma,\theta_{ext})
< 0.03
\left<F(E_{\gamma_0})\right>e^{\tau(E_{\gamma_0},z)}
\end{equation}  

Using Eq. (\ref{EB}) 
one finds that the magnetic
field strength probed by the study of  extended emission at $E_\gamma\simeq
0.5$~TeV at the angular scale $\theta_{ext}\simeq 0.5^\circ$
is $B \sim (3\div
6) \times 10^{-13}  \mbox{~G}$. 
Substituting (\ref{mkn3})
into (\ref{EB}) one finds
\begin{equation}  
{\cal V}\left[B\simeq (3\div 6)\times 10^{-13}\mbox{
G}\right]<0.4
\end{equation} 
which is much above the predictions of Ref. \cite{defl_Sigl} 
(${\cal V}(B\simeq 10^{-12}\mbox{~G})\sim 0.07$) and is just about the prediction of 
Ref. \cite{defl_Dolag} (${\cal V}(B\simeq 10^{-12}\mbox{~G})\sim 0.7$).  
The above example shows
that with modern telescopes, like HESS, MAGIC and VERITAS, one has a real 
possibility to detect or put tight
constraints on the EGMF in the $10^{-16}-10^{-12}$~G range if one makes a
systematic search of extended emission around a large number of point sources
 at different energies and different angular scales. 

Electromagnetic cascades of multi-TeV photons on EBL  were
considered  before in a different context. If the cascade happens
directly in or near the  source, secondary electrons and positrons
will be completely randomized in relatively  large magnetic fields,
$B>10^{-9}$~G, producing $R>1$~Mpc-size halo around the
source~\cite{Aharonian:halo}. In an opposite case, when EGMF is
extremely small, $B<10^{-18}$~G,  all the cascade will proceed  in the
forward direction and just contribute to the point source flux  (even
during the flares)~\cite{Plaga:flares,Aharonian:flares}. Contrary to
both those cases we are interested here in magnetic fields in the
intermediate range $10^{-16}$~G$<B<10^{-12}$~G,  in the voids of LSS,
and stress that such fields can be {\it measured} by detection of the
time independent  extended emission structures around observed TeV point sources.

To summarize, we have proposed a method of measurement of extra-galactic
magnetic fields with the help of TeV Cherenkov telescopes. The idea is to look
for extended emission produced by cascading of multi-TeV $\gamma$-rays emitted
by  distant point sources. The extended emission produced by the cascade in
relatively weak EGMF is expected to have a characteristic energy-dependent
surface brighness profile. Measuring the parameters of this profile enables one
to determine the characteristic of EGMF from Eqs. (\ref{EB}) and/or
(\ref{EB1}). The method is sensitive to EGMF strength in the range
$10^{-16}$~G$<B<10^{-12}$~G. 

We would like to thank Felix Aharonian, Igor Tkachev, Vladimir 
Vassiliev and Andrii Elyiv for fruitful comments to this manuscript.


\end{document}